\pdfoutput=1

\documentclass[11pt]{article}
\usepackage[final]{acl}

\usepackage{times}
\usepackage{latexsym}

\usepackage[T1]{fontenc}

\usepackage[utf8]{inputenc}

\usepackage{microtype}

\usepackage{inconsolata}

\usepackage{graphicx}
\usepackage{xspace}
\usepackage{booktabs}
\usepackage{amsmath}
\usepackage{amssymb}
\usepackage{mathtools}

\usepackage{algorithm}
\usepackage{algpseudocode}
\usepackage{algorithmicx}

\usepackage{makecell}

\usepackage{cleveref}

\usepackage{multirow} 

\usepackage{amsmath}

\usepackage{enumerate}
\usepackage{enumitem}
\usepackage{multirow}
\PassOptionsToPackage{table,xcdraw}{xcolor}
\usepackage{xcolor}

\usepackage[normalem]{ulem}
\useunder{\uline}{\ul}{}
\usepackage{pifont}
\usepackage{tcolorbox}

\newcommand{\cmark}{\ding{51}}

\newcommand{\modelname}{\textsc{PreMIR}\xspace}
\newcommand{\qclusername}{Q-Cluster\xspace}
\newcommand{\preq}{cross-modal preQs\xspace}

\newcommand{\github}{\raisebox{-1.5pt}{\includegraphics[height=1.05em]{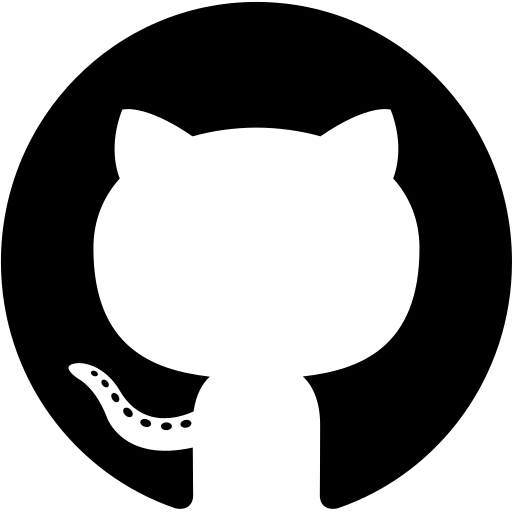}}\xspace}

\title{Zero-shot Multimodal Document Retrieval via\\Cross-modal Question Generation}

\newcommand*\samethanks[1][\value{footnote}]{\footnotemark[#1]}

\author{
Yejin Choi$^{\spadesuit}$\thanks{~Equal Contribution.} \quad
Jaewoo Park$^{\spadesuit}$\samethanks\\
\textbf{Janghan Yoon}$^{\spadesuit}$ \quad
\textbf{Saejin Kim}$^{\spadesuit}$ \quad
\textbf{Jaehyun Jeon}$^{\spadesuit}$ \quad
\textbf{Youngjae Yu}$^{\diamondsuit}$ \\
\small{$\spadesuit$ Yonsei University} \quad
\small{$\diamondsuit$ Seoul National University} \\
\texttt{\{yejinchoi, jerife\}@yonsei.ac.kr}\quad
\texttt{mycalljordan@snu.ac.kr}
}

\begin{document}
\maketitle

\begin{abstract}
Rapid advances in Multimodal Large Language Models (MLLMs) have extended information retrieval beyond text, enabling access to complex real-world documents that combine both textual and visual content. However, most documents are private, either owned by individuals or confined within corporate silos, and current retrievers struggle when faced with unseen domains or languages. To address this gap, we introduce \modelname, a simple yet effective framework that leverages the broad knowledge of an MLLM to generate cross-modal pre-questions (preQs) before retrieval. Unlike earlier multimodal retrievers that embed entire documents as a single vector, \modelname leverages preQs, decomposed from documents into finer token-level representations across modalities, enabling richer contextual understanding. Experiments show that \modelname achieves state-of-the-art performance on out-of-distribution benchmarks, including closed-domain and multilingual settings, outperforming strong baselines across all metrics. We confirm the contribution of each component through in-depth ablation studies, and qualitative analyses of the generated preQs further highlight the framework’s robustness in real-world settings\footnote{\github \textbf{Code}: \href{https://github.com/yejinc00/PREMIR}{yejinc00/PREMIR}}.
\end{abstract}
\section{Introduction}
\label{sec:introduction}

Advances in language models \cite{reimers-gurevych-2019-sentence} have enabled the creation of powerful retrievers that perform semantic search across documents, returning results closely aligned with user query \cite{karpukhin2020dense, khattab2020colbert}.  These retrievers are now widely deployed in real-world Retrieval-Augmented Generation (RAG) systems \cite{lewis2020retrieval}, where they assist Multimodal Large Language Models (MLLMs) \cite{xu2025qwen2, liu2023visual} by reducing hallucinations \cite{ayala2024reducing} and by supplying relevant context for evidence-guided answer generation \cite{jeong2024adaptive}.

In conventional RAG systems, chunk-based text retriever is widely adopted \cite{liu-etal-2024-lost}. However, this approach often overlooks crucial information such as images, tables, and document layout. Recent multimodal retrievers aim to address this limitation by extending retrieval capabilities to the visual domain, either by embedding both textual and visual elements using joint text–image encoders \cite{cao2019hybrid}, or by leveraging MLLMs to compute page-level embeddings directly \cite{faysse2024colpali, yu2024visrag}. Despite these advancements, current multimodal retrieval systems still encounter challenges in real-world scenarios, such as in personal workflows or corporate settings.

\begin{figure}[!t]
  \centering
  \includegraphics[width=1\linewidth]{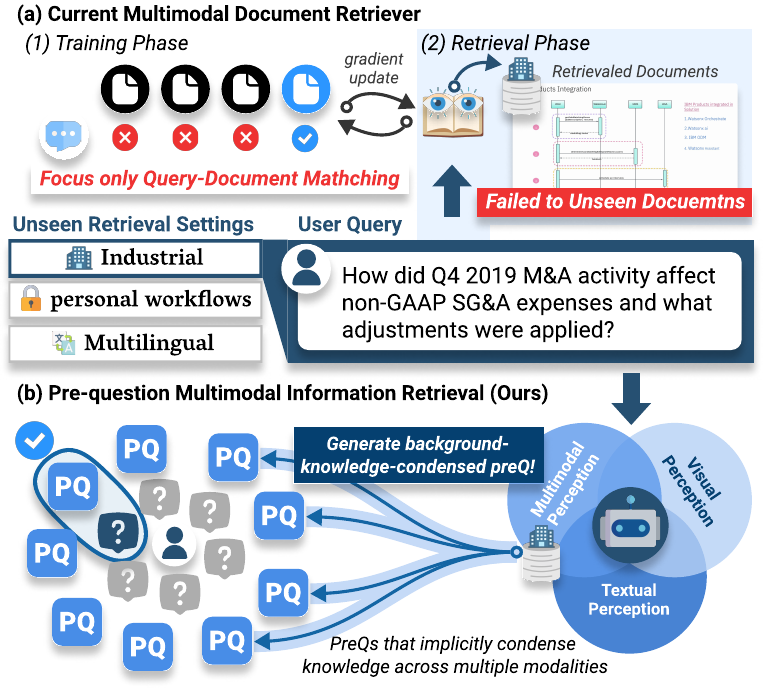}
    
    \vspace{-0.5em}
    \caption{In unseen documents retrieval settings, (a) conventional latent-level contrastive learning approaches in multimodal retrievers struggle to generalize. In contrast, (b) \modelname leverages token-level cross-modal complementary preQs to effectively handle such cases.}
  \label{fig:teaser}
\end{figure}

Multimodal retrievers exhibit significant performance degradation in \textit{out-of-distribution} settings, where documents contain content outside the scope of the training data. Existing multimodal models rely on directly comparing query embeddings with page image embeddings, typically encoding the entire images into vectors. This training objective allows distinguishment of relevant from irrelevant images through maximizing the similarity with positives and minimizing with negatives \cite{mnih2013learning, khattab2020colbert}. However, this training paradigm, focused on query-image alignment, often fails to learn a domain-transferable latent space, resulting in poor generalization to \textit{out-of-distribution} documents.

In addition, most existing methods treat MLLMs as static feature extractors, relying on fixed representations that overlook fine-grained cross-modal nuances. 
While some approaches \cite{nogueira2019doc2query, nogueira2019document, gospodinov2023doc2query} enrich document representations through query generation, they are limited to unimodal settings and remain highly dependent on the training distribution, further constraining their applicability in real-world scenarios.

To address these challenges, we propose the Pre-question Multimodal Information Retrieval (\textbf{\modelname}) framework, which generates cross-modal complementary pre-questions (preQs) from documents by leveraging the broad knowledge embedded in MLLMs, as shown in prior work to generalize across diverse domains \cite{yuan2023revisiting, alayrac2022flamingo, gruver2023large}. These cross-modal preQs inherently capture comprehensive background knowledge and diverse contextual information, ensuring robust and effective performance even in challenging \textit{out-of-distribution} scenarios, including multilingual and specialized closed-domain tasks. 
Furthermore, instead of embedding entire documents, the retriever compares queries, decomposed from documents into finer token-level representations across modalities, enabling richer contextual understanding.

Experimental results show that \modelname outperforms strong baselines on multimodal document retrieval in both closed-domain and multilingual settings, achieving state-of-the-art performance. 
Comprehensive ablation studies on each core module quantitatively confirm their individual contributions, and qualitative analyses offer intuitive insights into how our cross-modal preQs operate within the embedding space. 

In summary, our contributions are three-fold:

\begin{enumerate}[leftmargin=1em,topsep=0pt,itemsep=0ex,partopsep=1ex,parsep=1ex]
\item We propose \modelname, a multimodal retrieval framework that mitigates domain shift without training by generating cross-modal preQs.
\item \modelname achieves state-of-the-art performance on both multilingual and closed-domain benchmarks, showing strong real-world applicability.
\item Comprehensive ablation studies and analysis demonstrate how cross-modal preQs improve retrieval quality, offering insights into the mechanisms underlying \modelname's effectiveness.
\end{enumerate}
\begin{figure*}[t]
    \centering
    \includegraphics[width=1\linewidth]{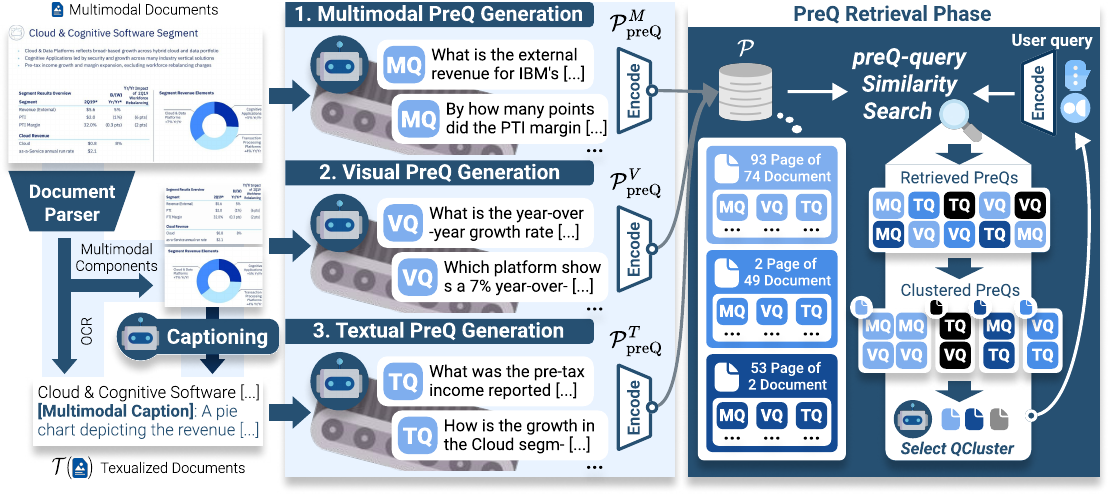}
    \vspace{-1.5em}
    \caption{Overview of the \modelname\ framework. \modelname\ first parses multimodal content in a modality-aware manner and generates multimodal, visual and textual preQs, which are stored in a shared embedding space. During retrieval, the preQs most similar to the user query are retrieved. In here, \qclusername module then clusters these preQs by their source passages and returns the clusters whose passages are contextually aligned with the user query.}
    \label{fig:preq_process}
\end{figure*}

\section{Method}
\modelname framework aims to generate cross-modal preQs that comprehensively cover the documents’ explicit and implicit knowledge from multimodal components, and retrieve the most appropriate semantically relevant preQs in response to a user query. In this section, we first outline the task definition in \Cref{subsec:task_def}, and then describe the cross-modal preQs generation and retrieval process of the \modelname framework in \Cref{subsec:preqir}.

\subsection{Task Definition}
\label{subsec:task_def}
\paragraph{Problem Setting.}
In multimodal RAG scenarios, several key design choices must be made. First is the choice of input modalities in the system - text, images, or both. Second is the level of granularity used for retrieval such as entire documents, individual pages, chunks, or specific image regions. Since real-world data is inherently multimodal and often distributed across heterogeneous sources, we adopt a practical \textit{out-of-distribution} configuration tailored for dynamic environments such as enterprise or personal workflows. In such settings, the corpus is typically domain-specific or multilingual, and the retrieval system must identify the most relevant passages (i.e., pages) from the entire document collection given an input text query.

\paragraph{Preliminary notations.}
We denote the text query as $q$, and the retriever searches for relevant passages $p_{i,j}$, where $p_{i,j}$ is the $j$-th passage (page) from the $i$-th document in the corpus $\mathcal{C} = \{\dots, p_{i,j}, \dots\}$. Each passage may contain text and multimodal components such as tables, figures, or charts. The retriever operates over a passage pool $\mathcal{P}$, which typically corresponds to the entire corpus $\mathcal{C}$.

\subsection{\modelname Framework}
\label{subsec:preqir}
\textit{Unlike} approaches that treat a page as a single image \cite{faysse2024colpali, yu2024visrag}, our framework captures the page along with fine-grained multimodal components, such as figures and OCR text regions within the page layout, to extract richer cross-modal features. As illustrated in \Cref{fig:preq_process}, we first parse every document to extract both visual and textual components, and then generate cross-modal preQs from this enriched representation to ensure diversity and contextual relevance. We employ a powerful MLLM, GPT-4o \cite{hurst2024gpt}, and prompts are provided in \Cref{sec:apdx_prompt}.

\paragraph{Multimodal Document Parsing.}
We employ a layout-aware document parser \cite{wang2024mineru} that fuses raw OCR output with grounded multimodal elements. For each page $p_{i,j}$, the parser returns the set of $k$ detected multimodal components (tables, figures, charts, and so on) denoted by $p_{i,j}^{\text{mc}}=\{mc_{1},\dots,mc_{k}\}$, and the OCR text $p_{i,j}^{\text{ocr}}$.

Next, every component in $p_{i,j}^{\text{mc}}$ is captioned with MLLM, and these captions are merged with $p_{i,j}^{\text{ocr}}$ while preserving the original layout order. The result is a layout-aware textual surrogate $p_{i,j}^{\text{text}}$ that faithfully reflects the multimodal content of the page. Finally, the triplet $\langle p_{i,j},p_{i,j}^{\text{mc}},p_{i,j}^{\text{text}}\rangle$, comprising the raw page image, its component images, and the textual surrogate, is passed downstream for cross-modal preQs generation.

\paragraph{Cross-modal PreQ Generation.}
Given a triplet $\langle p_{i,j},\,p_{i,j}^{\text{mc}},\,p_{i,j}^{\text{text}}\rangle$ for each page $(i,j)$ in the corpus $\mathcal{C}$, we construct three complementary preQ sets:
\begin{enumerate}[leftmargin=2em,topsep=0pt,itemsep=-1ex,partopsep=1ex,parsep=1ex,label=(\roman*)]
    \item \textit{Multimodal preQs}, $\mathcal{P}^{M}_{\text{preQ}}$, generated directly from the raw page image $p_{i,j}$ to preserve the original layout and cross-modal context;
    \item \textit{Visual preQs}, $\mathcal{P}^{V}_{\text{preQ}}$, created from individual visual components $p_{i,j}^{\text{mc}}$, such as figures, tables, and charts, to expose modality-specific cues;
    \item \textit{Textual preQs}, $\mathcal{P}^{T}_{\text{preQ}}$, derived from the layout-aware textual surrogate $p_{i,j}^{\text{text}}$.
\end{enumerate}

In conventional settings, the passage pool $\mathcal{P}$ from which the retriever selects candidate passages corresponds to the corpus $\mathcal{C}$.  
In contrast, we define the retrieval pool as the union of the three complementary preQ sets:
\begin{equation}
\mathcal{P}
   = \mathcal{P}^{M}_{\text{preQ}}
   \;\cup\;
     \mathcal{P}^{V}_{\text{preQ}}
   \;\cup\;
     \mathcal{P}^{T}_{\text{preQ}}.
\end{equation}
Each set $\mathcal{P}^\ast_{\text{preQ}}$ is generated by an MLLM \cite{hurst2024gpt}, which produces up to $n$ questions per passage in the corpus, with $n$ fixed at 50 in our experiments to balance performance and cost. These questions are designed to address not only information explicitly stated in the passage but also knowledge implicitly conveyed, as they are generated based on the broad knowledge of an MLLM.

\begin{table*}[!t]
\centering
\resizebox{0.98\linewidth}{!}{

\begin{tabular}{ll|cccc|cccc}
\toprule
                                                  &            & \multicolumn{4}{c|}{VIDoSeek}                                     & \multicolumn{4}{c}{REAL-MM-RAG}                                   \\ \midrule
                                                  & Model      & Recall@1       & Recall@3       & Recall@5       & MRR@5          & Recall@1       & Recall@3       & Recall@5       & MRR@5          \\ \midrule
                                                  & E5         & 0.488 & 0.715 & 0.802 & 0.611 & 0.176 & 0.280 & 0.328 & 0.228 \\
                                                  & GTE        & 0.415 & 0.617 & 0.715 & 0.528 & 0.175 & 0.276 & 0.320 & 0.229 \\
                                                  & BGE-M3     & 0.473 & 0.712 & 0.790 & 0.596 & 0.168 & 0.267 & 0.317 & 0.226 \\
\multirow{-4}{*}{\rotatebox[origin=c]{90}{Text}}  & ColBERT    & 0.556 & 0.744 & 0.819 & 0.656 & 0.171 & 0.261 & 0.305 & 0.220 \\ \midrule
                                                  & VisRAG-Ret & 0.638 & 0.843 & 0.911 & 0.746 & 0.282 & 0.438 & 0.502 & 0.365 \\
                                                  & ColPali    & 0.670 & 0.852 & 0.907 & 0.764 & 0.398 & 0.571 & 0.639 & 0.490 \\
\multirow{-3}{*}{\rotatebox[origin=c]{90}{Image}} & ColQwen2.0 & \underline{0.743} & \underline{0.912} & \underline{0.944} & \underline{0.827} & \underline{0.452} & \underline{0.622} & \underline{0.688} & \underline{0.543} \\ \midrule
\rowcolor[HTML]{ECF4FF} 
\multicolumn{2}{c|}{\cellcolor[HTML]{ECF4FF}\modelname (open)} & 0.690 & 0.861 & 0.900 & 0.777 & 0.437 & 0.598 & 0.643 & 0.520 \\ 
\rowcolor[HTML]{ECF4FF} 
\multicolumn{2}{c|}{\cellcolor[HTML]{ECF4FF}\modelname (closed)} & \textbf{0.797} & \textbf{0.918} & \textbf{0.952} & \textbf{0.861} & \textbf{0.500} & \textbf{0.673} & \textbf{0.724} & \textbf{0.589} \\ \bottomrule
\end{tabular}}

\caption{Experimental results for the zero-shot closed-domain, multimodal document retrieval task on VidoSeek \cite{wang2025vidorag} and REAL-MM-RAG \cite{wasserman2025real}. The best results are \textbf{boldfaced}, and the second-best results are \underline{underlined}.}
\label{tab:main}
\vspace{0.0em}
\end{table*}

\paragraph{\qclusername Retrieval.}
After constructing the retrieval pool $\mathcal{P}$, we embed each preQ using the retriever’s embedding.
Given a user query $q$, the retriever encodes it and retrieves the top-$k$ preQs from the retrieval pool $\mathcal{P}$ based on the highest cosine similarity of their embeddings.

A single preQ, however, may not fully capture the user’s intent, and enumerating every possible query would require an impractically large and diverse set.
To address this, we cluster preQs that were derived from the same source passage, thereby increasing the likelihood of retrieving a passage that directly answers the user’s query.

We first cluster the top-$k$ preQs originating from the same passage into a group $\mathcal{G}$. If the retrieval pool $\mathcal{P}$ contains more than 100$k$ entries, we set $k = 100$; otherwise, we set $k = 150$. Collecting all such groups yields $\mathcal{S} = \{\mathcal{G}_1, \dots, \mathcal{G}_m\}$. The LLM \cite{hurst2024gpt} then evaluates each cluster in $\mathcal{S}$ according to how well its associated passage answers the query and selects the most relevant candidates. By leveraging these preQ clusters, we alleviate the need to generate preQs that exhaustively cover all possible variations of user intent.
\begin{table*}[!t]
\centering
\resizebox{0.98\linewidth}{!}{

\begin{tabular}{l|cccc|cccc}
\toprule
           & \multicolumn{4}{c|}{$\text{CT}^2\text{C-QA}$ (Chinese)}   & \multicolumn{4}{c}{Allganize (Korean)} \\ \midrule
Model      & Recall@1 & Recall@3 & Recall@5 & MRR@5 & Recall@1 & Recall@3 & Recall@5 & MRR@5 \\ \midrule
ColBERT    & 0.048 & 0.097 & 0.132 & 0.077 & 0.056 & 0.107 & 0.125 & 0.082 \\
ColQwen2.0 & 0.126 & 0.228 & 0.295 & 0.185 & 0.565 & 0.748 & 0.813 & 0.659 \\ \midrule
\rowcolor[HTML]{ECF4FF}
\modelname (open) & \underline{0.255} & \textbf{0.405} & \textbf{0.477} & \textbf{0.337} & \underline{0.737} & \underline{0.863} & \underline{0.903} & \underline{0.805} \\
\rowcolor[HTML]{ECF4FF}
\modelname (closed) & \textbf{0.258} & \underline{0.399} & \underline{0.475} & \textbf{0.337} & \textbf{0.760} & \textbf{0.880} & \textbf{0.910} & \textbf{0.818} \\ \bottomrule
\end{tabular}}

\caption{Experimental results on the zero-shot multilingual, multimodal document-retrieval task for the Chinese benchmark $\text{CT}^{2}\text{C-QA}$ and the Korean benchmark Allganize RAG. Owing to its looser structure, $\text{CT}^{2}\text{C-QA}$ is markedly more challenging for baseline models than Allganize RAG.}
\label{tab:multilingual}
\vspace{0.0em}
\end{table*}

\section{Experiments}
To evaluate the robustness and adaptability of \modelname across different model scales, we conduct experiments with both closed-source and open-source models. For the open-source version, we include models of comparable size such as ColPaLI and ColQwen2.0, using Qwen3-Embedding-0.6B~\cite{zhang2025qwen3embeddingadvancingtext} for embedding and Qwen3-4B~\cite{yang2025qwen3technicalreport} for Q-clustering.

We evaluate \modelname under realistic multimodal retrieval conditions encompassing (i) multimodal inputs, (ii) multi-document collections, and (iii) closed-domain or multilingual scenarios, settings commonly encountered in both personal and industrial applications. We first describe the experimental setup in \cref{subsec:setup}. We then assess \modelname{} in closed-domain and multilingual environments, presented in \cref{subsec:exp_close} and \cref{subsec:exp_multilingual}, respectively. Additional details on the experimental setup and generated preQs are provided in \Cref{sec:apdx_Experiments_details}.

\subsection{Evaluation Settings}
\label{subsec:setup}

\paragraph{Baselines.}
Within the multimodal retrieval task defined in \cref{subsec:task_def}, we compare two categories of retrievers based on their input modality:

\textit{(1) Text-based.}  These models process passages only in textual form. To ensure a fair comparison, we provide the same parsed pages and VLM-generated captions introduced in \Cref{subsec:preqir}. The embedding-based retrievers included in our evaluation are E5~\cite{wang2022text}, GTE~\cite{li2023towards}, BGE-M3~\cite{chen2024bge}, and the late-interaction model ColBERT~\cite{khattab2020colbert}, which compute query and document token embeddings independently and match them.

\textit{(2) Image-based.}  In contrast to text-based, these models embed each document page as an image. VisRAG-Ret~\cite{yu2024visrag} leverages a MiniCPM-V~2.0~\cite{yao2024minicpm} + SigLIP~\cite{zhai2023sigmoid} MLLM backbone, whereas ColPaLI and ColQwen2~\cite{faysse2024colpali} adopt PaLI-3B~\cite{chen2022pali} and Qwen2-VL-2B~\cite{wang2024qwen2} backbones, respectively; both use the ColBERT scheme to match query–document pairs.

\paragraph{Metrics.}
We evaluate retrieval performance with two complementary metrics.  
Recall@$k$ measures coverage, the fraction of relevant passages that appear among the top-$k$ results; we report values for $k\!\in\!\{1,3,5\}$.  
MRR@$k$ captures how early the first relevant passage is retrieved, using $k\!=\!5$.  
Together, Recall@$k$ and MRR@5 reflect both breadth and ranking precision.

\subsection{Closed-domain Experiments}
\label{subsec:exp_close}
\paragraph{Setup.}
We evaluate two closed-domain benchmarks characterized by multimodal inputs and multi-document collections:
(i) ViDoSeek~\cite{wang2025vidorag} spans 12 topics, including economics, technology, literature, and geography, and contains 292 document decks with 5,385 passages and 1,142 queries, for which \modelname generates 328k preQs.
(ii) REAL-MM-RAG~\cite{wasserman2025real} targets industrial scenarios, providing 162 documents, a mixture of financial reports and technical manuals, yielding 8,604 passages, and 4,553 queries, for which \modelname produces 528k preQs.

\paragraph{Results.}
\Cref{tab:main} demonstrates that in closed-domain multimodal retrieval, the closed-\modelname outperforms all baselines on every benchmark without any additional multimodal retrieval training. The open-\modelname, while achieving relatively lower performance than the closed-setups, still surpasses ColPaLI. Text-based models often struggle to capture the distinctive features of multimodal inputs, leading to suboptimal performance. In contrast, image-based models generally perform better by overcoming some of these limitations; however, they still face challenges when handling unseen data in out-of-distribution documents scenarios. In our case, we leverage cross-modal preQs that implicitly condense knowledge across multiple modalities, enabling \modelname to generalize effectively to previously unseen data and achieve strong performance. These results demonstrate that \modelname generalizes robustly across both diverse personal topics and industrial corpora.

\subsection{Multilingual Experiments}
\label{subsec:exp_multilingual}

\paragraph{Setup.}
Following the closed-domain experiments reported in \cref{subsec:exp_close}, we use as baselines ColBERT and ColQwen 2.0, the highest-performing models for each input modality. We evaluate them on two public benchmarks:
(i) $\text{CT}^{2}\text{C-QA}$\cite{zhao2024ct2c} is a Chinese question-answering dataset compiled from the National Bureau of Statistics of China. Only a sampled subset is publicly available, consisting of 400 single-page passages and 20,480 queries. \modelname generates 58k preQs for this benchmark.
(ii) Allganize RAG\footnote{\url{https://huggingface.co/datasets/allganize/RAG-Evaluation-Dataset-KO}} is a Korean benchmark designed to evaluate RAG performance across domains such as finance, the public sector, healthcare, legal, and commerce. The publicly available dataset consists of 62 documents, resulting in 1289 passages and 278 queries. \modelname produces 56k preQs for this dataset.

\paragraph{Results.}
\Cref{tab:multilingual} shows that \modelname consistently outperforms all baselines across every dataset and metric in the multilingual setting. On the Chinese benchmark, the documents are loosely curated, creating a more realistic retrieval scenario in which both text- and image-based models struggle. Even under these conditions, \modelname surpasses the strong baseline ColQwen2.0 by more than a factor of two in Recall@1. For the Korean benchmark, whose passages are comparatively well organized, ColQwen2.0 attains higher scores than it does on the Chinese; however, its performance still drops in the closed, multilingual context, whereas \modelname maintains a clear lead. These findings imply that \modelname generalizes robustly in multilingual closed-domain retrieval, reinforcing its suitability for real-world applications.
\section{Analysis of \modelname}
\subsection{Ablation Study}
To investigate the effectiveness of \modelname's core modules and to justify our design choices, we conduct ablation experiments in this section.

\paragraph{Retrieval Ablation.} To analyze whether our approach of generating and retrieving preQs performs better than conventional page-level retrieval, we fix the embedding model and conduct both approaches. As shown in \Cref{tab:ablation_retrieval}, our approach consistently outperforms across all metrics, demonstrating that the improvement stems not merely from using a stronger embedding model but from the effectiveness of our approach itself.

\paragraph{Cross-modal PreQ Ablation.} The results in \Cref{tab:ablation_preq} show that combining all three PreQ types achieves the best performance across all metrics. In particular, using the full set, yields the highest Recall@1 and MRR@5 scores, indicating that the three types complement each other. When used individually, $\mathcal{P}^{M}_{\text{preQ}}$ substantially outperforms both $\mathcal{P}^{V}_{\text{preQ}}$ and $\mathcal{P}^{T}_{\text{preQ}}$, underscoring the importance of preserving the original layout and cross‑modal context in document understanding tasks.

\begin{table}[!t]
\centering
\resizebox{1\linewidth}{!}{

\begin{tabular}{l|ccc}
\toprule
Retrieval Type     & Recall@1 & Recall@3 & MRR@5 \\ \midrule
\rowcolor[HTML]{ECF4FF} 
preQs (\modelname) & \textbf{0.678}    & \textbf{0.916}    & \textbf{0.77}  \\
Texts (Conventional)              & 0.630     & 0.845    & 0.739\\ 
\bottomrule
\end{tabular}}

\caption{Ablation results on VideoSeek without preQ clustering, comparing preQ-based retrieval with conventional text-based retrieval.}
\label{tab:ablation_retrieval}
\vspace{0.0em}
\end{table}
\begin{table}[!t]
\centering
\resizebox{1\linewidth}{!}{
\begin{tabular}{ccc|ccc}
\toprule

$\mathcal{P}^M_\text{preQ}$ & $\mathcal{P}^V_\text{preQ}$ & $\mathcal{P}^T_\text{preQ}$ & Recall@1       & Recall@5       & MRR@5          \\ \midrule
\rowcolor[HTML]{ECF4FF} 
\cmark                      & \cmark                      & \cmark                      & \textbf{0.678} & 0.916          & \textbf{0.770} \\
\cmark                      & \cmark                      & \multicolumn{1}{l|}{}       & 0.672          & \textbf{0.919} & 0.770          \\
\cmark                      & \multicolumn{1}{l}{}        & \cmark                      & 0.672          & 0.909          & 0.764          \\
\multicolumn{1}{l}{}        & \cmark                      & \cmark                      & 0.590          & 0.869          & 0.701          \\
\cmark                      & \multicolumn{1}{l}{}        & \multicolumn{1}{l|}{}       & 0.652          & 0.913          & 0.755          \\
\multicolumn{1}{l}{}        & \cmark                      & \multicolumn{1}{l|}{}       & 0.397          & 0.588          & 0.471          \\
\multicolumn{1}{l}{}        & \multicolumn{1}{l}{}        & \cmark                      & 0.568          & 0.858          & 0.680

\\ \bottomrule
\end{tabular}}
\caption{Ablation study results of multimodal preQs $\mathcal{P}^{M}_{\text{preQ}}$, visual preQs $\mathcal{P}^{V}_{\text{preQ}}$, and textual preQs $\mathcal{P}^{T}_{\text{preQ}}$ on the VidoSeek without preQ clustering.}
\label{tab:ablation_preq}
\vspace{0.0em}
\end{table}

\paragraph{\qclusername Ablation.} Table~\ref{tab:ablation_module} highlights the substantial performance gains obtained by introducing our \qclusername mechanism. Specifically, \qclusername improves Recall@1 by 0.119, Recall@5 by 0.036, and MRR@5 by 0.091. This lightweight module helps the system prioritize passages that better address the query, confirming its value in retrieval.

To assess its practicality, we replace \qclusername's backbone LLM with alternatives and report results in \Cref{tab:ablation_llm}. \modelname delivers consistent performance across all language models. While GPT-4o~\cite{hurst2024gpt} achieves the best scores, open‑weight models such as DeepSeek‑V3\cite{liu2024deepseek}, Qwen2.5‑72B~\cite{bai2025qwen2}, Llama3.3‑72B~\cite{grattafiori2024llama}, and even the compact Qwen2.5‑7B suffer only minor degradation. These findings indicate that \modelname effectively leverages open-weight models to achieve state-of-the-art multimodal document retrieval.

\begin{table}[!t]
\centering
\resizebox{0.82\linewidth}{!}{
\begin{tabular}{l|ccc}
\toprule

Model               & Recall@1       & Recall@5       & MRR@5          \\ \midrule
\rowcolor[HTML]{ECF4FF} 
\modelname             & \textbf{0.797} & \textbf{0.952} & \textbf{0.861} \\
\hspace{2pt}- Qcluster & 0.678          & 0.916          & 0.770

\\ \bottomrule
\end{tabular}}
\caption{Ablation study results for the process of clustering the retrieved preQs and selecting the cluster that best satisfies the query over the VidoSeek.}
\label{tab:ablation_module}
\vspace{0.0em}
\end{table}

\begin{table}[!t]
\centering
\resizebox{0.9\linewidth}{!}{
\begin{tabular}{l|ccc}
\toprule

Model           & Recall@1       & Recall@5       & MRR@5          \\ \midrule
\rowcolor[HTML]{ECF4FF} 
GPT-4o          & \textbf{0.797} & \textbf{0.952} & \textbf{0.861} \\
DeepSeek-V3     & 0.758          & 0.943          & 0.837          \\
Qwen2.5$_{\text{72B}}$   & 0.762          & 0.933          & 0.834          \\
Llama-3.3$_{\text{72B}}$ & 0.751          & 0.941          & 0.828          \\
Qwen2.5$_{\text{7B}}$    & 0.736          & 0.928          & 0.813

\\ \bottomrule
\end{tabular}}
\caption{Impact of different LLMs in the \qclusername module on retrieval performance over the VidoSeek.}
\label{tab:ablation_llm}
\vspace{0.0em}
\end{table}
\begin{table}[!t]
\centering
\resizebox{1\linewidth}{!}{
\begin{tabular}{l|ccc}
\toprule

Model                  & Recall@1       & Recall@5       & MRR@5          \\ \midrule
\rowcolor[HTML]{ECF4FF} 
text-embedding-3-large & \textbf{0.678} & \textbf{0.916} & \textbf{0.770} \\
bge-large-en-v1.5      & 0.603          & 0.886          & 0.713          \\
gte-Qwen2-7B-instruct         & 0.576          & 0.878          & 0.691

\\ \bottomrule
\end{tabular}}
\caption{Comparison of retrieval performance using different embedding backbones on VideoSeek without PreQ clustering.}
\label{tab:ablation_emb}
\vspace{0.0em}
\end{table}

\paragraph{Embedding Model Ablation.}
\Cref{tab:ablation_emb} shows the results obtained with two open-weight embedding models, BGE \cite{chen2024bge} and the Qwen2-based GTE \cite{li2023towards}. \modelname delivers competitive retrieval quality even with these fully open embeddings, eliminating the need for proprietary solutions. Although the closed-weight baseline attains the highest overall score, the open-weight BGE and GTE variants remain close, especially in Recall@5, where they reach 0.886 and 0.878, respectively, versus 0.916. This narrow gap demonstrates that \modelname maintains robust retrieval capability with widely accessible embeddings, making it practical for diverse deployment scenarios.

\begin{figure}[!t]
  \centering
  \includegraphics[width=1\linewidth]{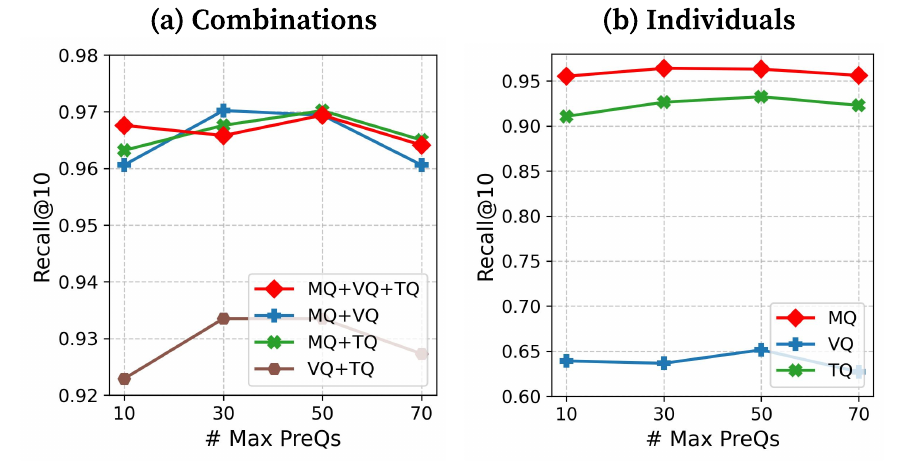}
    \caption{Ablation study on varying the number of generated preQs in ViDoSeek. (a) The left side shows results using combined multimodal, visual, and textual PreQs, while the (b) right side shows results using each modality individually.}
  \label{fig:analysis_n}
  \vspace{-0.0em}
\end{figure}

\subsection{Cross-modal PreQ Analysis}

\paragraph{Impact of the Number of PreQs.}
\Cref{fig:analysis_n} shows that retrieval performance varies only marginally with different numbers of preQs ($n$), indicating limited gains from simply increasing $n$. Yet benchmark recall alone may underestimate practical utility, as it only partially reflects real-world query diversity. To address this, we analyze semantic cluster formation as $n$ grows, as shown in \Cref{tab:analysis_preq}. Specifically, we embed the preQs using Qwen3-Embedding \cite{yang2025qwen3technicalreport} and apply DBSCAN clustering \cite{10.5555/3001460.3001507}, which automatically determines the number of clusters. We find that recall quickly plateaus, whereas cluster diversity continues to increase until convergence. This suggests that although a small $n$ (e.g., 10) suffices for benchmarks, adaptively selecting $n$ based on semantic coverage offers a more principled strategy for real-world scenarios.

\begin{table}[!t]
\centering
\resizebox{0.8\linewidth}{!}{

\begin{tabular}{c|cc}
\toprule
         & \multicolumn{2}{c}{Avg. \# of generated clusters} \\ \midrule
\# of PreQs & VIDoSeek               & REAL-MM-RAG              \\ \midrule
10       & 16.02                  & 14.48                    \\
30       & 26.53                  & 21.36                    \\
50       & 29.79                  & 22.47                    \\
70       & 31.19                  & 23.77                    \\
\bottomrule
\end{tabular}}

\caption{Analysis of how the number of preQs influences the formation of semantic clusters on benchmarks. While Recall plateaus quickly (\Cref{fig:analysis_n}), but clusters keep increasing, indicating broader coverage.}
\label{tab:analysis_preq}
\vspace{0.0em}
\end{table}

\paragraph{Quality Analysis of PreQs.}
To assess the quality of generated PreQs, we analyze both redundancy and specificity.
\Cref{tab:analysis_cos} reports redundancy using cosine similarity on ViDoSeek. With our redundancy-reducing prompt design, only 0.6\% of PreQs from the same page exceeded a similarity of 0.9, and across documents only 0.21\% exceeded 0.6, confirming that redundancy is effectively minimized.
For specificity, we conduct an LLM-based annotation (1–5 Likert scale \cite{zheng2023judging}) on a 10\% sample of ViDoSeek (in \Cref{tab:analysis_likert}). Only about 13\% of PreQs were rated as generic (scores 1–2), indicating that most were highly domain-specific. Moreover, such generic PreQs were retrieved only ~8\%, showing that our pipeline remains robust to generic questions.

\begin{table}[!t]
\centering
\resizebox{0.94\linewidth}{!}{

\begin{tabular}{c|cc}
\toprule
\makecell[l]{Similarity\\Threshold} & 
\makecell{\% of similar pairs\\within same source} & 
\makecell{\% of similar pairs\\across all PreQs} \\ \midrule
$\geq$ 0.5                             & 57.96                              & 1.92                             \\
$\geq$ 0.6                             & 35.73                               & 0.21                               \\
$\geq$ 0.7                             & 17.47                               & 0.02                               \\
$\geq$ 0.8                             & 5.36                                 & 0.00                               \\
$\geq$ 0.9                             & 0.67                                 & 0.00                              \\
\bottomrule
\end{tabular}}

\caption{Analysis of cosine similarity between preQ pairs generated within the same document and across all preQs, showing minimal redundancy.}
\label{tab:analysis_cos}
\vspace{0.0em}
\end{table}
\begin{table}[!t]
\centering
\resizebox{0.82\linewidth}{!}{

\begin{tabular}{c|cc}
\toprule
\makecell{Likert\\scale}        & \makecell{\% across all\\generated PreQs} & \makecell{\% among\\retrieved PreQs} \\ \midrule
1    & 10.10                        & 6.80                    \\
2         & 3.35                         & 1.98                    \\
3        & 37.84                        & 27.69                   \\
4        & 29.51                        & 31.03                   \\
5 & 19.19                        & 32.51              \\
\bottomrule
\end{tabular}}

\caption{Analysis of PreQ specificity using a 1–5 Likert scale where 1 indicates generic and 5 indicates specific, showing that most PreQs are highly domain-specific.}
\label{tab:analysis_likert}
\vspace{0.0em}
\end{table}

\begin{figure}[!t]
  \centering
  \includegraphics[width=1\linewidth]{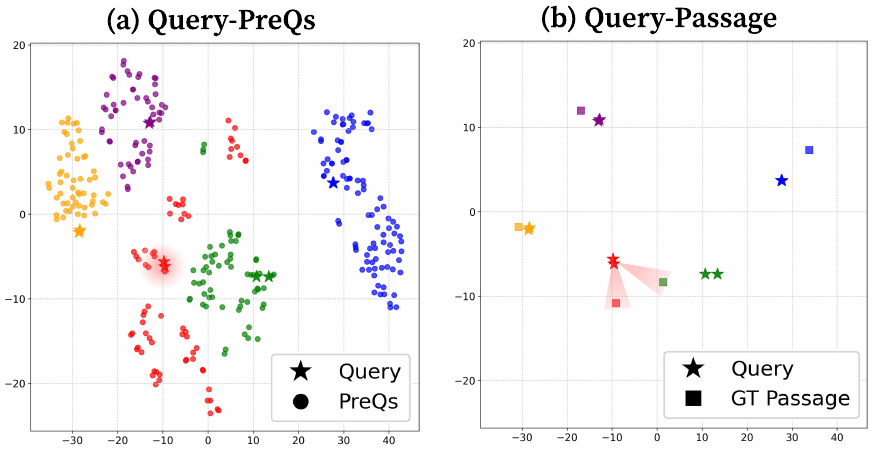}
    \caption{Comparison of query to preQ retrieval and query to passage retrieval. Objects of the same color represent the ground truth retrieval targets.}
  \label{fig:analysis_comparison}
  \vspace{0.0em}
\end{figure}

\begin{figure*}[t]
    \centering
    \includegraphics[width=1\linewidth]{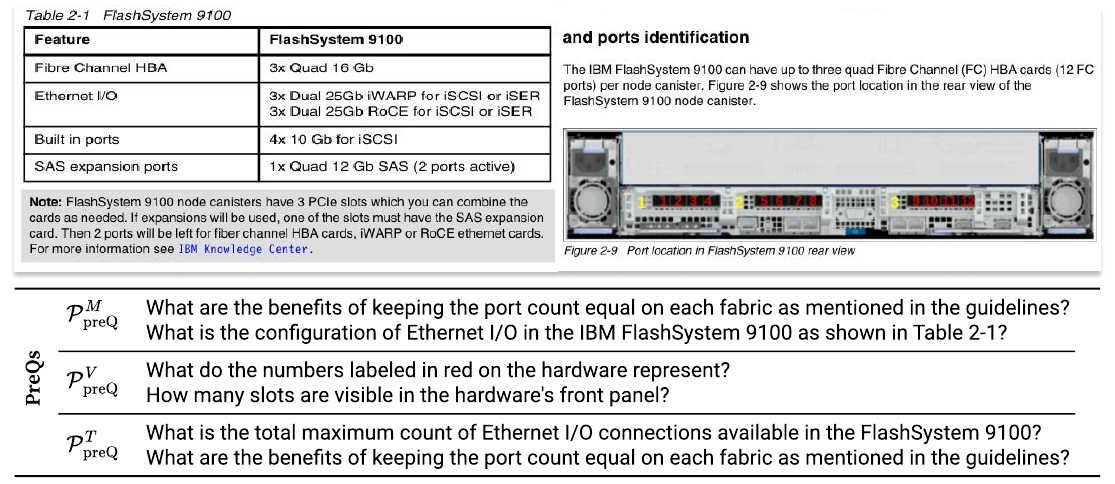}
    \caption{Qualitative examples of multimodal, visual, and textual preQs generated from the passage above. The multimodal preQs capture the overall context of the document, while the visual and textual preQ focus on specific visual and linguistic details, respectively.}
    \label{fig:qualitative_results}
    \vspace{-1em}
\end{figure*}

\begin{figure}[!t]
  \centering
  \includegraphics[width=1.0\linewidth]{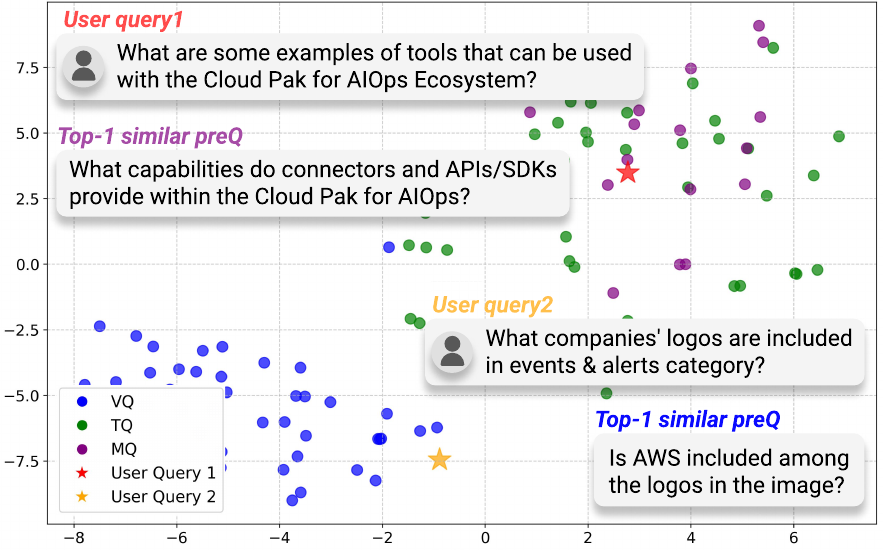}
    \caption{User query and \preq in the embedding space visualized with t-SNE \cite{JMLR:v9:vandermaaten08a}. The top-1 multimodal and visual PreQs are well aligned with the user’s intent.}
  \label{fig:analysis_triplet}
  \vspace{0.0em}
\end{figure}

\paragraph{Improved Passage Discrimination.}
\Cref{fig:analysis_comparison} compares conventional query–passage retrieval with the \modelname by examining the embedding‐space distances between a user query and candidate passages. In conventional retrieval, embeddings of incorrect passages often lie close to the query as well as to the correct target passage, making misretrievals more likely, an especially critical issue when the pool of relevant passages is small. By contrast, \modelname alleviates this problem, its use of cross-modal preQs generates intermediate representations that carve out clearer semantic boundaries between passage clusters. As a result, embeddings of correct targets are more cleanly separated from those of confusable passages, leading to more reliable discrimination during retrieval.

\paragraph{Synergy of Cross-modal PreQs.}
\label{subsec:exp_qualitative}
As illustrated in \Cref{fig:qualitative_results,fig:analysis_triplet}, multimodal, visual, and textual preQs form a complementary triad that broadens document coverage and embedding-space reach.

At the document level, multimodal preQs $(\mathcal{P}^{M}_{\text{preQ}})$ analyze the document holistically, integrating content, tables, and visual elements to generate questions focused on overall narrative flow and high-level semantics. Visual preQs $(\mathcal{P}^{V}_{\text{preQ}})$ specifically process image inputs, generating targeted questions tailored to visual content without encompassing the document's broader context. Textual preQs $(\mathcal{P}^{T}_{\text{preQ}})$ delve deeply into fine-grained linguistic aspects, such as entity mentions and definitions, providing detailed linguistic context.

In the embedding space, these complementary modalities enhance retrieval accuracy across diverse query types by occupying distinct regions. For instance, as demonstrated with \texttt{user query2}, which emphasizes specific visual elements, visual PreQs $(\mathcal{P}^{V}_{\text{preQ}})$ effectively address such queries by leveraging modality-specific features embedded within figures, and other visual components. This strategy ensures comprehensive document understanding and consistently improves retrieval performance across diverse queries.

\begin{table}[!t]
\centering
\resizebox{1\linewidth}{!}{

\begin{tabular}{l|ccc|cc}
\toprule

                                           & \multicolumn{3}{c|}{Offline (page/s)} & \multicolumn{2}{c}{Online  (query/s)} \\ \midrule
Model                                      & Parse      & Q-Gen      & Index     & Retrieve           & Cluster           \\ \midrule
ColBERT                                    & 5.10         & -             & 0.01        & 0.01                 & -                   \\
ColQwen2.0                                 & -            & -             & 1.30        & 0.34                 & -                   \\ \midrule
\modelname                                 & 5.10         & 16.42         & 33.94       & 0.56                 & 0.82                \\
\rowcolor[HTML]{ECF4FF} 
\rotatebox[origin=c]{180}{$\Lsh$} Optimized & 0.51         & 0.90          & 0.08        & 0.22                 & 0.02

\\ \bottomrule
\end{tabular}}

\caption{Latency analysis of offline (page/s) and online (query/s) phases across models.}
\label{tab:analysis_time}
\vspace{0.0em}
\end{table}

\subsection{Applicability of \modelname}
To evaluate the applicability of \modelname, we conduct both latency and cost analyses.
\Cref{tab:analysis_time} reports offline and online latency for the standard and optimized versions, with detailed settings provided in the \Cref{sec:apdx_applicability_setting}. Under optimized settings, \modelname achieves 0.1 seconds lower online latency than ColQwen2.0. Although slower than ColBERT, it substantially outperforms ColBERT in retrieval performance.
We also estimate computational cost and show in the \Cref{sec:apdx_applicability_cost} that \modelname is more cost-efficient than competing models.
\section{Related Work}

\subsection{Multimodal Document Retrieval}
Recent efforts to bridge the semantic gap between queries and documents have explored diverse approaches. Early dense retrievers such as DPR~\cite{karpukhin2020dense} and ColBERT~\cite{khattab2020colbert} improved text matching, while multimodal models like LayoutLM~\cite{xu2020layoutlm}, DocFormer~\cite{appalaraju2021docformer}, and UDOP~\cite{tang2023unifying} advanced the joint use of textual, visual, and layout features. More recent systems, including ColPaLI~\cite{faysse2024colpali}, ColQwen2, and VisRAG-Ret~\cite{yu2024visrag}, leverage multimodal large language models (MLLMs) such as MiniCPM-V 2.0~\cite{yao2024minicpm} and Qwen2-VL~\cite{Qwen2VL} to encode documents as images and compare them with query embeddings. However, these contrastive learning–based approaches remain vulnerable to unseen queries and out-of-distribution (OOD) documents. To address this limitation, \modelname leverages the prior knowledge of MLLMs to generate multimodal preQs that naturally incorporate OOD information, enabling token-level matching. This approach achieves stronger performance on OOD benchmarks without additional training.

\subsection{Applications of Query Expansion}
Query expansion techniques have been applied to address challenges across various domains. In dialogue systems, expanding conversational queries with contextual information \cite{ni2023recent} enhances coherence and response quality. For domain-specific search, query expansion has bridged terminology gaps in medicine \cite{peikos2024leveraging} and law \cite{nguyen2024enhancing}. To address vocabulary mismatch in information retrieval, Doc2query \cite{nogueira2019document} pioneered predicting potential queries, later refined by DocT5query with T5's pre-trained knowledge, while InPars \cite{bonifacio2022inpars} leveraged LLMs for synthetic query generation. However, these methods remain limited to text and fail to capture cross-modal interactions critical for multimodal document retrieval, whereas \modelname overcomes these limitations by leveraging multimodal preQs, enabling comprehensive cross-modal understanding and robust retrieval performance.

\section{Conclusion}
We introduced \modelname, a powerful multimodal retrieval framework utilizing the broad knowledge of a MLLM to generate cross-modal preQs prior to retrieval. Unlike traditional multimodal retrieval methods limited by distribution-dependent training, our proposed cross-modal preQs implicitly condense information across modalities, enabling strong out-of-distribution retrieval performance. Remarkably, \modelname achieves state-of-the-art results across all metrics under challenging out-of-distribution scenarios, including closed-domain and multilingual settings, without requiring additional training. Comprehensive ablation studies and analysis further demonstrate the effectiveness of cross-modal preQs in significantly enhancing retrieval quality, providing insights into the underlying mechanisms and highlighting the strong potential of \modelname for real-world applications.

\section{Limitations}
\modelname shows a limitation in consistently generating specific cross-modal PreQs using an MLLM. Despite explicit instructions, the model occasionally produces generic questions due to the subjective nature of `specificity'. Fortunately, these generic PreQs have minimal impact on retrieval performance, as they are less likely to match user queries and rank low. Future work should focus on enhancing specificity, either by suppressing generic questions during generation or applying filtering mechanisms. Additionally, adaptive PreQ generation based on document complexity may improve efficiency by reducing computational costs.


\bibliography{custom}

\begin{thebibliography}{47}
\providecommand{\natexlab}[1]{#1}

\bibitem[{Alayrac et~al.(2022)Alayrac, Donahue, Luc, Miech, Barr, Hasson, Lenc, Mensch, Millican, Reynolds et~al.}]{alayrac2022flamingo}
Jean-Baptiste Alayrac, Jeff Donahue, Pauline Luc, Antoine Miech, Iain Barr, Yana Hasson, Karel Lenc, Arthur Mensch, Katherine Millican, Malcolm Reynolds, and 1 others. 2022.
\newblock Flamingo: a visual language model for few-shot learning.
\newblock \emph{Advances in neural information processing systems}, 35:23716--23736.

\bibitem[{Appalaraju et~al.(2021)Appalaraju, Jasani, Kota, Xie, and Manmatha}]{appalaraju2021docformer}
Srikar Appalaraju, Bhavan Jasani, Bhargava~Urala Kota, Yusheng Xie, and R~Manmatha. 2021.
\newblock Docformer: End-to-end transformer for document understanding.
\newblock In \emph{Proceedings of the IEEE/CVF international conference on computer vision}, pages 993--1003.

\bibitem[{Ayala and Bechard(2024)}]{ayala2024reducing}
Orlando Ayala and Patrice Bechard. 2024.
\newblock Reducing hallucination in structured outputs via retrieval-augmented generation.
\newblock In \emph{Proceedings of the 2024 Conference of the North American Chapter of the Association for Computational Linguistics: Human Language Technologies (Volume 6: Industry Track)}, pages 228--238.

\bibitem[{Bai et~al.(2025)Bai, Chen, Liu, Wang, Ge, Song, Dang, Wang, Wang, Tang et~al.}]{bai2025qwen2}
Shuai Bai, Keqin Chen, Xuejing Liu, Jialin Wang, Wenbin Ge, Sibo Song, Kai Dang, Peng Wang, Shijie Wang, Jun Tang, and 1 others. 2025.
\newblock Qwen2. 5-vl technical report.
\newblock \emph{arXiv preprint arXiv:2502.13923}.

\bibitem[{Bonifacio et~al.(2022)Bonifacio, Abonizio, Fadaee, and Nogueira}]{bonifacio2022inpars}
Luiz Bonifacio, Hugo Abonizio, Marzieh Fadaee, and Rodrigo Nogueira. 2022.
\newblock Inpars: Unsupervised dataset generation for information retrieval.
\newblock In \emph{Proceedings of the 45th International ACM SIGIR Conference on Research and Development in Information Retrieval}, pages 2387--2392.

\bibitem[{Cao et~al.(2019)Cao, Lin, He, and He}]{cao2019hybrid}
Wenming Cao, Qiubin Lin, Zhihai He, and Zhiquan He. 2019.
\newblock Hybrid representation learning for cross-modal retrieval.
\newblock \emph{Neurocomputing}, 345:45--57.

\bibitem[{Chen et~al.(2024)Chen, Xiao, Zhang, Luo, Lian, and Liu}]{chen2024bge}
Jianlv Chen, Shitao Xiao, Peitian Zhang, Kun Luo, Defu Lian, and Zheng Liu. 2024.
\newblock Bge m3-embedding: Multi-lingual, multi-functionality, multi-granularity text embeddings through self-knowledge distillation.
\newblock \emph{arXiv preprint arXiv:2402.03216}.

\bibitem[{Chen et~al.(2022)Chen, Wang, Changpinyo, Piergiovanni, Padlewski, Salz, Goodman, Grycner, Mustafa, Beyer et~al.}]{chen2022pali}
Xi~Chen, Xiao Wang, Soravit Changpinyo, AJ~Piergiovanni, Piotr Padlewski, Daniel Salz, Sebastian Goodman, Adam Grycner, Basil Mustafa, Lucas Beyer, and 1 others. 2022.
\newblock Pali: A jointly-scaled multilingual language-image model.
\newblock \emph{arXiv preprint arXiv:2209.06794}.

\bibitem[{Ester et~al.(1996)Ester, Kriegel, Sander, and Xu}]{10.5555/3001460.3001507}
Martin Ester, Hans-Peter Kriegel, J\"{o}rg Sander, and Xiaowei Xu. 1996.
\newblock A density-based algorithm for discovering clusters in large spatial databases with noise.
\newblock In \emph{Proceedings of the Second International Conference on Knowledge Discovery and Data Mining}, KDD'96, page 226–231. AAAI Press.

\bibitem[{Faysse et~al.(2024)Faysse, Sibille, Wu, Omrani, Viaud, Hudelot, and Colombo}]{faysse2024colpali}
Manuel Faysse, Hugues Sibille, Tony Wu, Bilel Omrani, Gautier Viaud, C{\'e}line Hudelot, and Pierre Colombo. 2024.
\newblock Colpali: Efficient document retrieval with vision language models.
\newblock In \emph{The Thirteenth International Conference on Learning Representations}.

\bibitem[{Gospodinov et~al.(2023)Gospodinov, MacAvaney, and Macdonald}]{gospodinov2023doc2query}
Mitko Gospodinov, Sean MacAvaney, and Craig Macdonald. 2023.
\newblock Doc2query--: when less is more.
\newblock In \emph{European Conference on Information Retrieval}, pages 414--422. Springer.

\bibitem[{Grattafiori et~al.(2024)Grattafiori, Dubey, Jauhri, Pandey, Kadian, Al-Dahle, Letman, Mathur, Schelten, Vaughan et~al.}]{grattafiori2024llama}
Aaron Grattafiori, Abhimanyu Dubey, Abhinav Jauhri, Abhinav Pandey, Abhishek Kadian, Ahmad Al-Dahle, Aiesha Letman, Akhil Mathur, Alan Schelten, Alex Vaughan, and 1 others. 2024.
\newblock The llama 3 herd of models.
\newblock \emph{arXiv preprint arXiv:2407.21783}.

\bibitem[{Gruver et~al.(2023)Gruver, Finzi, Qiu, and Wilson}]{gruver2023large}
Nate Gruver, Marc Finzi, Shikai Qiu, and Andrew~G Wilson. 2023.
\newblock Large language models are zero-shot time series forecasters.
\newblock \emph{Advances in Neural Information Processing Systems}, 36:19622--19635.

\bibitem[{Hurst et~al.(2024)Hurst, Lerer, Goucher, Perelman, Ramesh, Clark, Ostrow, Welihinda, Hayes, Radford et~al.}]{hurst2024gpt}
Aaron Hurst, Adam Lerer, Adam~P Goucher, Adam Perelman, Aditya Ramesh, Aidan Clark, AJ~Ostrow, Akila Welihinda, Alan Hayes, Alec Radford, and 1 others. 2024.
\newblock Gpt-4o system card.
\newblock \emph{arXiv preprint arXiv:2410.21276}.

\bibitem[{Jeong et~al.(2024)Jeong, Baek, Cho, Hwang, and Park}]{jeong2024adaptive}
Soyeong Jeong, Jinheon Baek, Sukmin Cho, Sung~Ju Hwang, and Jong~C Park. 2024.
\newblock Adaptive-rag: Learning to adapt retrieval-augmented large language models through question complexity.
\newblock In \emph{Proceedings of the 2024 Conference of the North American Chapter of the Association for Computational Linguistics: Human Language Technologies (Volume 1: Long Papers)}, pages 7029--7043.

\bibitem[{Karpukhin et~al.(2020)Karpukhin, O{\u{g}}uz, Min, Lewis, Wu, Edunov, Chen, and Yih}]{karpukhin2020dense}
Vladimir Karpukhin, Barlas O{\u{g}}uz, Sewon Min, Patrick Lewis, Ledell Wu, Sergey Edunov, Danqi Chen, and Wen-tau Yih. 2020.
\newblock Dense passage retrieval for open-domain question answering.
\newblock \emph{arXiv preprint arXiv:2004.04906}.

\bibitem[{Khattab and Zaharia(2020)}]{khattab2020colbert}
Omar Khattab and Matei Zaharia. 2020.
\newblock Colbert: Efficient and effective passage search via contextualized late interaction over bert.
\newblock In \emph{Proceedings of the 43rd International ACM SIGIR conference on research and development in Information Retrieval}, pages 39--48.

\bibitem[{Lewis et~al.(2020)Lewis, Perez, Piktus, Petroni, Karpukhin, Goyal, K{\"u}ttler, Lewis, Yih, Rockt{\"a}schel et~al.}]{lewis2020retrieval}
Patrick Lewis, Ethan Perez, Aleksandra Piktus, Fabio Petroni, Vladimir Karpukhin, Naman Goyal, Heinrich K{\"u}ttler, Mike Lewis, Wen-tau Yih, Tim Rockt{\"a}schel, and 1 others. 2020.
\newblock Retrieval-augmented generation for knowledge-intensive nlp tasks.
\newblock \emph{Advances in Neural Information Processing Systems}, 33:9459--9474.

\bibitem[{Li et~al.(2023)Li, Zhang, Zhang, Long, Xie, and Zhang}]{li2023towards}
Zehan Li, Xin Zhang, Yanzhao Zhang, Dingkun Long, Pengjun Xie, and Meishan Zhang. 2023.
\newblock Towards general text embeddings with multi-stage contrastive learning.
\newblock \emph{arXiv preprint arXiv:2308.03281}.

\bibitem[{Liu et~al.(2024{\natexlab{a}})Liu, Feng, Xue, Wang, Wu, Lu, Zhao, Deng, Zhang, Ruan et~al.}]{liu2024deepseek}
Aixin Liu, Bei Feng, Bing Xue, Bingxuan Wang, Bochao Wu, Chengda Lu, Chenggang Zhao, Chengqi Deng, Chenyu Zhang, Chong Ruan, and 1 others. 2024{\natexlab{a}}.
\newblock Deepseek-v3 technical report.
\newblock \emph{arXiv preprint arXiv:2412.19437}.

\bibitem[{Liu et~al.(2023)Liu, Li, Wu, and Lee}]{liu2023visual}
Haotian Liu, Chunyuan Li, Qingyang Wu, and Yong~Jae Lee. 2023.
\newblock Visual instruction tuning.
\newblock \emph{Advances in neural information processing systems}, 36:34892--34916.

\bibitem[{Liu et~al.(2024{\natexlab{b}})Liu, Lin, Hewitt, Paranjape, Bevilacqua, Petroni, and Liang}]{liu-etal-2024-lost}
Nelson~F. Liu, Kevin Lin, John Hewitt, Ashwin Paranjape, Michele Bevilacqua, Fabio Petroni, and Percy Liang. 2024{\natexlab{b}}.
\newblock \href {https://doi.org/10.1162/tacl_a_00638} {Lost in the middle: How language models use long contexts}.
\newblock \emph{Transactions of the Association for Computational Linguistics}, 12:157--173.

\bibitem[{Mnih and Kavukcuoglu(2013)}]{mnih2013learning}
Andriy Mnih and Koray Kavukcuoglu. 2013.
\newblock Learning word embeddings efficiently with noise-contrastive estimation.
\newblock \emph{Advances in neural information processing systems}, 26.

\bibitem[{Nguyen et~al.(2024)Nguyen, Nguyen, Nguyen, Nguyen, Vuong, and Satoh}]{nguyen2024enhancing}
Hai-Long Nguyen, Duc-Minh Nguyen, Tan-Minh Nguyen, Ha-Thanh Nguyen, Thi-Hai-Yen Vuong, and Ken Satoh. 2024.
\newblock Enhancing legal document retrieval: A multi-phase approach with large language models.
\newblock \emph{arXiv preprint arXiv:2403.18093}.

\bibitem[{Ni et~al.(2023)Ni, Young, Pandelea, Xue, and Cambria}]{ni2023recent}
Jinjie Ni, Tom Young, Vlad Pandelea, Fuzhao Xue, and Erik Cambria. 2023.
\newblock Recent advances in deep learning based dialogue systems: A systematic survey.
\newblock \emph{Artificial intelligence review}, 56(4):3055--3155.

\bibitem[{Nogueira et~al.(2019{\natexlab{a}})Nogueira, Lin, and Epistemic}]{nogueira2019doc2query}
Rodrigo Nogueira, Jimmy Lin, and AI~Epistemic. 2019{\natexlab{a}}.
\newblock From doc2query to doctttttquery.
\newblock \emph{Online preprint}, 6(2).

\bibitem[{Nogueira et~al.(2019{\natexlab{b}})Nogueira, Yang, Lin, and Cho}]{nogueira2019document}
Rodrigo Nogueira, Wei Yang, Jimmy Lin, and Kyunghyun Cho. 2019{\natexlab{b}}.
\newblock Document expansion by query prediction.
\newblock \emph{arXiv preprint arXiv:1904.08375}.

\bibitem[{Peikos et~al.(2024)Peikos, Kasela, and Pasi}]{peikos2024leveraging}
Georgios Peikos, Pranav Kasela, and Gabriella Pasi. 2024.
\newblock Leveraging large language models for medical information extraction and query generation.
\newblock \emph{arXiv preprint arXiv:2410.23851}.

\bibitem[{{Qwen Team}(2025)}]{yang2025qwen3technicalreport}
{Qwen Team}. 2025.
\newblock \href {https://arxiv.org/abs/2505.09388} {Qwen3 technical report}.
\newblock \emph{Preprint}, arXiv:2505.09388.

\bibitem[{Reimers and Gurevych(2019)}]{reimers-gurevych-2019-sentence}
Nils Reimers and Iryna Gurevych. 2019.
\newblock \href {https://doi.org/10.18653/v1/D19-1410} {Sentence-{BERT}: Sentence embeddings using {S}iamese {BERT}-networks}.
\newblock In \emph{Proceedings of the 2019 Conference on Empirical Methods in Natural Language Processing and the 9th International Joint Conference on Natural Language Processing (EMNLP-IJCNLP)}, pages 3982--3992, Hong Kong, China. Association for Computational Linguistics.

\bibitem[{Tang et~al.(2023)Tang, Yang, Wang, Fang, Liu, Zhu, Zeng, Zhang, and Bansal}]{tang2023unifying}
Zineng Tang, Ziyi Yang, Guoxin Wang, Yuwei Fang, Yang Liu, Chenguang Zhu, Michael Zeng, Cha Zhang, and Mohit Bansal. 2023.
\newblock Unifying vision, text, and layout for universal document processing.
\newblock In \emph{Proceedings of the IEEE/CVF conference on computer vision and pattern recognition}, pages 19254--19264.

\bibitem[{van~der Maaten and Hinton(2008)}]{JMLR:v9:vandermaaten08a}
Laurens van~der Maaten and Geoffrey Hinton. 2008.
\newblock \href {http://jmlr.org/papers/v9/vandermaaten08a.html} {Visualizing data using t-sne}.
\newblock \emph{Journal of Machine Learning Research}, 9(86):2579--2605.

\bibitem[{Wang et~al.(2024{\natexlab{a}})Wang, Xu, Zhao, Ouyang, Wu, Zhao, Xu, Liu, Qu, Shang et~al.}]{wang2024mineru}
Bin Wang, Chao Xu, Xiaomeng Zhao, Linke Ouyang, Fan Wu, Zhiyuan Zhao, Rui Xu, Kaiwen Liu, Yuan Qu, Fukai Shang, and 1 others. 2024{\natexlab{a}}.
\newblock Mineru: An open-source solution for precise document content extraction.
\newblock \emph{arXiv preprint arXiv:2409.18839}.

\bibitem[{Wang et~al.(2022)Wang, Yang, Huang, Jiao, Yang, Jiang, Majumder, and Wei}]{wang2022text}
Liang Wang, Nan Yang, Xiaolong Huang, Binxing Jiao, Linjun Yang, Daxin Jiang, Rangan Majumder, and Furu Wei. 2022.
\newblock Text embeddings by weakly-supervised contrastive pre-training.
\newblock \emph{arXiv preprint arXiv:2212.03533}.

\bibitem[{Wang et~al.(2024{\natexlab{b}})Wang, Bai, Tan, Wang, Fan, Bai, Chen, Liu, Wang, Ge, Fan, Dang, Du, Ren, Men, Liu, Zhou, Zhou, and Lin}]{Qwen2VL}
Peng Wang, Shuai Bai, Sinan Tan, Shijie Wang, Zhihao Fan, Jinze Bai, Keqin Chen, Xuejing Liu, Jialin Wang, Wenbin Ge, Yang Fan, Kai Dang, Mengfei Du, Xuancheng Ren, Rui Men, Dayiheng Liu, Chang Zhou, Jingren Zhou, and Junyang Lin. 2024{\natexlab{b}}.
\newblock Qwen2-vl: Enhancing vision-language model's perception of the world at any resolution.
\newblock \emph{arXiv preprint arXiv:2409.12191}.

\bibitem[{Wang et~al.(2024{\natexlab{c}})Wang, Bai, Tan, Wang, Fan, Bai, Chen, Liu, Wang, Ge et~al.}]{wang2024qwen2}
Peng Wang, Shuai Bai, Sinan Tan, Shijie Wang, Zhihao Fan, Jinze Bai, Keqin Chen, Xuejing Liu, Jialin Wang, Wenbin Ge, and 1 others. 2024{\natexlab{c}}.
\newblock Qwen2-vl: Enhancing vision-language model's perception of the world at any resolution.
\newblock \emph{arXiv preprint arXiv:2409.12191}.

\bibitem[{Wang et~al.(2025)Wang, Ding, Chen, Wu, Wang, Xie, and Zhao}]{wang2025vidorag}
Qiuchen Wang, Ruixue Ding, Zehui Chen, Weiqi Wu, Shihang Wang, Pengjun Xie, and Feng Zhao. 2025.
\newblock Vidorag: Visual document retrieval-augmented generation via dynamic iterative reasoning agents.
\newblock \emph{arXiv preprint arXiv:2502.18017}.

\bibitem[{Wasserman et~al.(2025)Wasserman, Pony, Naparstek, Goldfarb, Schwartz, Barzelay, and Karlinsky}]{wasserman2025real}
Navve Wasserman, Roi Pony, Oshri Naparstek, Adi~Raz Goldfarb, Eli Schwartz, Udi Barzelay, and Leonid Karlinsky. 2025.
\newblock Real-mm-rag: A real-world multi-modal retrieval benchmark.
\newblock \emph{arXiv preprint arXiv:2502.12342}.

\bibitem[{Xu et~al.(2025)Xu, Guo, He, Hu, He, Bai, Chen, Wang, Fan, Dang et~al.}]{xu2025qwen2}
Jin Xu, Zhifang Guo, Jinzheng He, Hangrui Hu, Ting He, Shuai Bai, Keqin Chen, Jialin Wang, Yang Fan, Kai Dang, and 1 others. 2025.
\newblock Qwen2. 5-omni technical report.
\newblock \emph{arXiv preprint arXiv:2503.20215}.

\bibitem[{Xu et~al.(2020)Xu, Li, Cui, Huang, Wei, and Zhou}]{xu2020layoutlm}
Yiheng Xu, Minghao Li, Lei Cui, Shaohan Huang, Furu Wei, and Ming Zhou. 2020.
\newblock Layoutlm: Pre-training of text and layout for document image understanding.
\newblock In \emph{Proceedings of the 26th ACM SIGKDD international conference on knowledge discovery \& data mining}, pages 1192--1200.

\bibitem[{Yao et~al.(2024)Yao, Yu, Zhang, Wang, Cui, Zhu, Cai, Li, Zhao, He et~al.}]{yao2024minicpm}
Yuan Yao, Tianyu Yu, Ao~Zhang, Chongyi Wang, Junbo Cui, Hongji Zhu, Tianchi Cai, Haoyu Li, Weilin Zhao, Zhihui He, and 1 others. 2024.
\newblock Minicpm-v: A gpt-4v level mllm on your phone.
\newblock \emph{arXiv preprint arXiv:2408.01800}.

\bibitem[{Yu et~al.(2024)Yu, Tang, Xu, Cui, Ran, Yan, Liu, Wang, Han, Liu et~al.}]{yu2024visrag}
Shi Yu, Chaoyue Tang, Bokai Xu, Junbo Cui, Junhao Ran, Yukun Yan, Zhenghao Liu, Shuo Wang, Xu~Han, Zhiyuan Liu, and 1 others. 2024.
\newblock Visrag: Vision-based retrieval-augmented generation on multi-modality documents.
\newblock \emph{arXiv preprint arXiv:2410.10594}.

\bibitem[{Yuan et~al.(2023)Yuan, Chen, Cui, Gao, Zou, Cheng, Ji, Liu, and Sun}]{yuan2023revisiting}
Lifan Yuan, Yangyi Chen, Ganqu Cui, Hongcheng Gao, Fangyuan Zou, Xingyi Cheng, Heng Ji, Zhiyuan Liu, and Maosong Sun. 2023.
\newblock Revisiting out-of-distribution robustness in nlp: Benchmarks, analysis, and llms evaluations.
\newblock \emph{Advances in Neural Information Processing Systems}, 36:58478--58507.

\bibitem[{Zhai et~al.(2023)Zhai, Mustafa, Kolesnikov, and Beyer}]{zhai2023sigmoid}
Xiaohua Zhai, Basil Mustafa, Alexander Kolesnikov, and Lucas Beyer. 2023.
\newblock Sigmoid loss for language image pre-training.
\newblock In \emph{Proceedings of the IEEE/CVF international conference on computer vision}, pages 11975--11986.

\bibitem[{Zhang et~al.(2025)Zhang, Li, Long, Zhang, Lin, Yang, Xie, Yang, Liu, Lin, Huang, and Zhou}]{zhang2025qwen3embeddingadvancingtext}
Yanzhao Zhang, Mingxin Li, Dingkun Long, Xin Zhang, Huan Lin, Baosong Yang, Pengjun Xie, An~Yang, Dayiheng Liu, Junyang Lin, Fei Huang, and Jingren Zhou. 2025.
\newblock \href {https://arxiv.org/abs/2506.05176} {Qwen3 embedding: Advancing text embedding and reranking through foundation models}.
\newblock \emph{Preprint}, arXiv:2506.05176.

\bibitem[{Zhao et~al.(2024)Zhao, Cheng, Zhang, Cheng, Feng, and Zhang}]{zhao2024ct2c}
Bowen Zhao, Tianhao Cheng, Yuejie Zhang, Ying Cheng, Rui Feng, and Xiaobo Zhang. 2024.
\newblock Ct2c-qa: Multimodal question answering over chinese text, table and chart.
\newblock In \emph{Proceedings of the 32nd ACM International Conference on Multimedia}, pages 3897--3906.

\bibitem[{Zheng et~al.(2023)Zheng, Chiang, Sheng, Zhuang, Wu, Zhuang, Lin, Li, Li, Xing et~al.}]{zheng2023judging}
Lianmin Zheng, Wei-Lin Chiang, Ying Sheng, Siyuan Zhuang, Zhanghao Wu, Yonghao Zhuang, Zi~Lin, Zhuohan Li, Dacheng Li, Eric Xing, and 1 others. 2023.
\newblock Judging llm-as-a-judge with mt-bench and chatbot arena.
\newblock \emph{Advances in neural information processing systems}, 36:46595--46623.

\end{thebibliography}

\clearpage
\appendix
\section{Implementation details}
\label{sec:apdx_implementation_details}
\subsection{Benchmark Details}
Since the $\text{CT}^{2}\text{C-QA}$ dataset was not officially available, we utilized only 400 samples. For the Allganize dataset, we constructed our dataset by selecting only the documents that are practically available.
The distribution of multimodal, visual, and textual pre-questions across the datasets used in this study is summarized in Table~\ref{tab:tab_apdx_preq_dist}.

\subsection{\modelname details}
For retrieval, we used OpenAI's text embedding model \textit{text-embedding-3-large} for query embedding. The LLM used in Qcluster is \textit{gpt-4o}. MQ and VQ generation was done using \textit{gpt-4o}, while TQ generation was done using \textit{gpt-4o-mini}.
For captioning the parsed components, we used \textit{gpt-4o-mini}.

\subsection{Experimental Details}
\label{sec:apdx_Experiments_details}
All experiments were run with three random seeds, and the standard deviations across runs were below 0.01. Experiments for VisRAG-Ret, ColQwen2.0, and ColPali were conducted on an NVIDIA RTX 3090 GPU.

\section{Limitations details}
\begin{figure}[!t]
  \centering
  \includegraphics[width=0.48\textwidth]{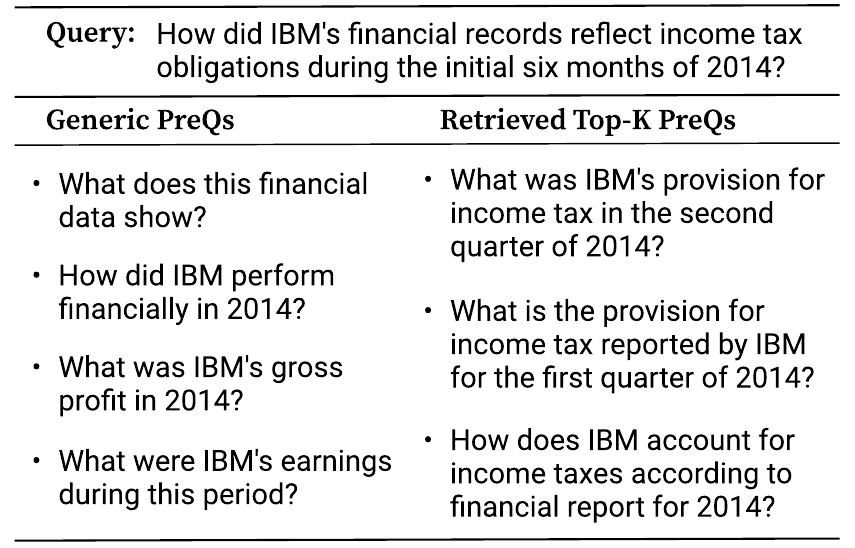}
    \caption{Limitation example of generated PreQs.}
  \label{fig:analysis_limitation}
  \vspace{0.0em}
\end{figure}

MLLM occasionally generates generic questions alongside specific ones. While these generic PreQs could potentially lead to incorrect retrieval, Figure ~\ref{fig:analysis_limitation} demonstrates that they rarely appear in top-K results when users submit specific queries. The retrieval process naturally filters out generic PreQs as they lack the distinctive characteristics to match well with specific user information needs. Therefore, despite the challenge of generating consistently specific PreQs, generic questions have minimal impact on overall system performance.

\section{Prompt}
\label{sec:apdx_prompt}
This section presents the prompts used throughout the image parsing and question generation pipeline. For image captioning, we refer to the prompt in \Cref{fig:prompt_apdx_image_caption}. For pre-question generation, we show the prompts used for multimodal pre-questions, visual pre-questions (\Cref{fig:prompt_apdx_preq_image}), and textual pre-questions (\Cref{fig:prompt_apdx_preq_text}). Additionally, the prompt used for \qclusername\ is also provided in \Cref{fig:prompt_apdx_qcluster}.

\begin{table}[!t]
\centering
\resizebox{0.48\textwidth}{!}{
\begin{tabular}{lrrrr}
\toprule

Dataset & MQ & VQ & TQ & Total \\
\midrule
VidoSeek   & 49,523  & 46,659  & 232,003  & 328,185 \\
REAL-MM-RAG    & 107,137 & 90,433  & 384,352  & 581,922 \\
$\text{CT}^2\text{C-QA}$      & 8,976   & 31,523  & 17,418   & 57,917 \\
Allganize  & 8,979   & 7,625   & 39,177   & 55,781

\\ \bottomrule
\end{tabular}}
\caption{Statistics of question types in each dataset: MQ (multimodal preQs), VQ (visual preQs), and TQ (textual preQs).}
\label{tab:tab_apdx_preq_dist}
\vspace{0.0em}
\end{table}

\section{Details of Applicability Analysis}
\subsection{Optimized Setting Details of \modelname}
\label{sec:apdx_applicability_setting}

For the standard setting, ColBERT and ColQwen2.0 were run on an AMD EPYC 9354 32-core CPU with an RTX 4090 GPU, while \modelname was executed on an AMD EPYC 7413 CPU at 1.71GHz. For the optimized setting:

\begin{itemize}[leftmargin=1em,topsep=0pt,itemsep=0ex,partopsep=1ex,parsep=1ex]
    \item Query \& PreQ Embedding: AMD EPYC 7413 CPU @ 1.71GHz, asynchronous multi-threading with 60 threads, batch size = 200.
    \item PreQ Retrieval: AMD EPYC 7413 CPU @ 1.71GHz, asynchronous multi-processing with 96 processes, batch size = total\_queries / worker\_count.
    \item Q-Clustering: AMD EPYC 7413 CPU @ 1.71GHz, asynchronous multi-threading with 60 threads.
\end{itemize}

\subsection{Cost Analysis of PreQ Generation}
\label{sec:apdx_applicability_cost}
We evaluated the computational cost of PreQ generation on the ViDoSeek benchmark (5,385 pages). Using the open-source Qwen2.5VL-72B model on 8 $\times$ RTX 3090 GPUs, generation required over 1,400 GPU hours, corresponding to approximately \$5,481 on AWS g4dn.12xlarge instances. A smaller 7B model under the same setting reduced the cost to about \$217.  

In contrast, a proprietary API was substantially more efficient, completing generation in 24.6 hours at a total cost of \$40.9, which could be further reduced to 1.3 hours and \$20.5 with multiprocessing and batching. Restricting generation to MQ alone reduced the cost even further, to roughly \$9, with little impact on performance. These results highlight that efficient configurations make large-scale PreQ generation practical and scalable.

\section{Icon Attribution}
The icons used in the figures were obtained from Flaticon \url{https://www.flaticon.com} and are attributed to their respective authors in accordance with Flaticon's license.

\begin{figure*}[t]
\centering
\begin{tcolorbox}[
    colback=white, %
    colframe=gray, %
    arc=4mm, %
    fontupper=\small %
]
You are given an image that represents part of a document, such as a figure, table, chart, or diagram.\\

Your task is to generate a clear, informative, and self-contained caption that describes: \\
1. What kind of image this is (e.g., chart, table, photograph, infographic) — provide a high-level description.\\
2. The detailed content within the image, including specific values, trends, comparisons, categories, or key insights, if applicable.\\

If the image contains a data visualization (e.g., a chart or table), describe the type of data, major trends, significant differences, or any notable patterns.\\

Avoid referring to the image as "this image" or using phrases like "shown here." Just write the caption as if it were placed directly below the image.
\end{tcolorbox}
\caption{A prompt for generating captioned images during document parsing. The inputs of the prompts are \textbf{boldfaced} and image.}
\label{fig:prompt_apdx_image_caption}
\end{figure*}
\begin{figure*}[t]
\centering
\begin{tcolorbox}[
    colback=white, %
    colframe=gray, %
    arc=4mm, %
    fontupper=\small %
]
You are a helpful assistant for generating pre-questions based on a document.\\

Your task is to create "pre-questions" that a user might naturally ask **before** reading the document.\\

Each pre-question must satisfy the following conditions: \\

1. The question must be **specific and clearly formulated**, since it is asked \textit{before} reading the document. \\
\hspace*{1em}-- Do **not** use vague expressions like "this model", "in this document", or "According to the table".\\
\hspace*{1em}-- Instead, **explicitly mention** the target of the question. \\
\hspace*{1em}-- For example: "What is the performance of model A on dataset B?"\\

2. The question must have a **clear and verifiable answer within the document itself**.\\
\hspace*{1em}-- Do not generate questions that cannot be answered using the document's content.\\

3. Generate up to \textbf{\{cfg.max\_new\_questions\}} questions.\\
\hspace*{1em}-- All questions must be **diverse and non-redundant**.\\
\hspace*{1em}-- Avoid repeating the same type of question or asking the same thing in different ways.\\

**Output format**:\\
-- Return the questions as a JSON array of objects.\\
-- Each object must follow this format: \\

\{\{ \\
\hspace*{1em}"question": "string" \\
\}\} \\

\vspace{1em}
**- - -** \\
**Document**: \\
\{document\_text\} \\
**- - -** \\
**Output**:
\end{tcolorbox}
\caption{A prompt designed to create both visual and multimodal pre-questions. The inputs of the prompts are \textbf{boldfaced}.}
\label{fig:prompt_apdx_preq_text}
\end{figure*}
\begin{figure*}[t]
\centering
\begin{tcolorbox}[
    colback=white, %
    colframe=gray, %
    arc=4mm, %
    fontupper=\small %
]
You are a helpful assistant for generating pre-questions based on an image-based document.\\

Your task is to create "pre-questions" that a user might naturally ask **before** reading this image-based document.\\

Each pre-question must satisfy the following conditions: \\

1. The question must be **specific and clearly formulated**, since it is asked \textit{before} reading the document.\\
\hspace*{1em}-- Do **not** use vague expressions like "this model", "in this document", or "According to the table".\\
\hspace*{1em}-- Instead, **explicitly mention** the target of the question.\\
\hspace*{1em}-- For example: "What is the performance of model A on dataset B?"\\

2. The question must have a **clear and verifiable answer within the document itself**.\\
\hspace*{1em}-- The answer should be grounded in the document's content, including **multimodal elements** such as:\\
\hspace*{2em}- Figures (e.g., line graphs, bar charts)\\
\hspace*{2em}- Tables with numerical or categorical data\\
\hspace*{2em}- Diagrams, labeled illustrations, or structured visual layouts\\
\hspace*{1em}-- Do not generate questions that cannot be answered using these visual or textual components.\\

3. Generate up to \textbf{\{cfg.max\_new\_questions\}} questions.\\
\hspace*{1em}-- All questions must be **diverse and non-redundant**.\\
\hspace*{1em}-- Avoid repeating the same type of question or asking the same thing in different ways.\\

**Output format**:\\
-- Return the questions as a JSON array of objects.\\
-- If the document contains no visual elements, return an empty list: \texttt{[]}\\
-- Otherwise, format your output as a JSON array, where each object has the following structure:\\

\begin{verbatim}
[
  {
    "question": "string"
  }
]
\end{verbatim}

**- - -**\\
**Output**:
\end{tcolorbox}
\caption{A prompt designed to create both visual and multimodal preQs. The inputs of the prompts are \textbf{boldfaced} and image.}
\label{fig:prompt_apdx_preq_image}
\end{figure*}
\begin{figure*}[t]
\centering
\begin{tcolorbox}[
    colback=white, %
    colframe=gray, %
    arc=4mm, %
    fontupper=\small %
]
User query: \textbf{\{query\}}\\
Retrieved questions (grouped by source):\\
\textbf{\{questions\_text\}}\\

Each question belongs to a source group (e.g., same document or generator). Some questions may be semantically similar because they come from the same source.\\

Please rank the TOP 5 source groups by how relevant and helpful their associated questions are for answering the user's query. Within each group, consider the best representative question to assess relevance.\\

Your goal is to select and rank the top 5 most useful groups such that the most useful ones are listed first, based on semantic similarity to the user's query.\\

IMPORTANT: Only include the 5 MOST RELEVANT group numbers in your ranking. If there are fewer than 5 groups total, include all of them.\\

Output only the group numbers in ranked order, separated by commas.\\
Example output: 2,1,4,3,5
\end{tcolorbox}
\caption{A prompt used for \qclusername. The inputs of the prompts are \textbf{boldfaced}.}
\label{fig:prompt_apdx_qcluster}
\end{figure*}

\end{document}